# Exchange bias in Co-$Cr_2O_3$ nanocomposites

P Anil Kumar[a)] and K Mandal[b)]

Magnetism Laboratory, S N Bose National Centre for Basic Sciences,

Block JD, Sector III, Salt Lake, Kolkata-700 098

**Abstract**

The possibility of using exchange bias in ferromagnetic-antiferromagnetic system to over come the effect of superparamagnetism in small cobalt nanoparticles is explored. We have prepared Co-$Cr_2O_3$ nanocomposite powders using a chemical method. It is shown that in this system the effect of superparamagnetism in cobalt nanoparticles could be overcome. The superparamagnetic blocking temperature of 3 nm cobalt particles has been increased to above room temperature. The choice of $Cr_2O_3$ is vital as its $T_N$ is higher compared to other antiferromagnetic materials used for this purpose such as CoO. The field cooled and zero field cooled hysteresis measurements of the samples confirm the existence of exchange bias interaction in this system.

[a)] anil@bose.res.in

[b)] kalyan@bose.res.in

## I. Introduction:

The possibility of achieving ultra high-density data storage system using magnetic particles as memory elements and their uses in medicine has recently enhanced the research activities in the area of magnetic nanomaterials [1-3]. These applications need the smallest possible magnetic particles for effective usage. However, as the size of the magnetic grain is reduced below a critical size, the domain wall formation is not supported energetically and hence the single magnetic domain particles are developed. These single domain particles are of great physical interest as they show strikingly new phenomena [4-6]. On further reduction of grain size, the anisotropy energy (being proportional to particle volume) holding the particle magnetization in a particular direction becomes comparable to that of the thermal energy at and above superaparamagnetic blocking temperature $T_B^{SP}$. This makes the particle magnetization to flip in all easy directions. As a result of it; the individual particles lose their ferromagnetic behavior. This phenomenon, called superparamagnetism [7,8] limits the use of ultra fine particles for various applications.

To increase the thermal stability of ultra fine particles by using ferromagnetic (FM)-antiferromagnetic (AFM) exchange bias [9-11] system is an important topic of research in magnetism in the direction of ultra high density storage media. The existence of spin-spin exchange interaction at the interface of FM and AFM materials can qualitatively explain the existence of extra anisotropy energy in this (FM-AFM) system. This extra anisotropy energy turns out to be unidirectional favoring the alignment of FM magnetization in a single direction. When a large field is applied in the temperature range $T_N<T<T_C$ ($T_N$, $T_C$ are the Neel and Curie temperature of AFM and FM material respectively), the FM spins line up with the field, while the AFM spins remain random. On cooling the sample to $T<T_N$ keeping the magnetic field on, the AFM spins next to the FM spins, align ferromagnetically to those of the FM (assuming ferromagnetic interaction at the interface). When the field is reversed, the FM spins start to rotate. However, for sufficiently large AFM anisotropy, the AFM spins remain unchanged. Therefore, the interfacial interaction between the FM/AFM spins at the interface, tries to align the FM spins ferromagnetically with the AFM spins at the interface. In other words, the AFM spins at the interface exert a microscopic torque on the FM spins, to keep them in their original position (ferromagnetically aligned at the interface). Therefore, the FM spins have one single stable configuration, i.e. the anisotropy is *unidirectional*. Thus, the field needed to reverse completely an FM layer magnetization will be larger if it is in contact with an AFM, because an extra field is needed to overcome the microscopic torque of AFM spins on FM spins. However, once the field is rotated back to its original direction, the FM spins will start to rotate at a smaller field, due to the interaction with the AFM spins (which now exert a torque in the same direction as the external field). The material behaves as if there was an extra (internal) biasing field; therefore, the FM hysteresis loop is shifted in the field axis, i.e. *exchange bias*. Since this exchange bias phenomenon is unidirectional it can be used to successfully lock the magnetization of superparamagnetic particles in a single direction against to the thermal flipping. Hence this leads to the overcoming of superparamagnetic limit. However, the Neel temperature of AFM phase limits the temperature up to which exchange bias effect exists, leading to the definition of another blocking temperature $T_B^{EB}$. At this point we would like to mention that the microscopic origin of the exchange bias is open for the debate [12, 13].

Though most studied are multi-layer systems; the polycrystalline powder systems do show considerable exchange bias effect [14]. In fact the phenomenon of exchange bias was first observed by Meiklejohn and Bean in Co-CoO particulate system [10]. Skumryev et al [9] have used the exchange bias effect to overcome the superparamagnetic limit in Co-CoO multilayers. They have enhanced the superparamagnetic blocking temperature of 4nm cobalt nanoparticles from 10K to around 290K. In the present work we have studied the effect of exchange bias on the superparamagnetism of cobalt nanoparticles embedded in $Cr_2O_3$ matrix. The choice of $Cr_2O_3$ facilitates the use of the exchange bias system to above 300K. This leads to the non-zero hysteresis of fine cobalt particles at room temperature.

## II. Experimental:

The Co-$Cr_2O_3$ nanocomposite powders were prepared by thermal decomposition of the combined solution of the metal nitrates. The calculated amounts of cobalt nitrate [$Co(NO_3)_2$. $6H_2O$] and chromium nitrate [$Cr(NO_3)_3$. $9H_2O$] were taken so as to have the desired weight percentage of Co and $Cr_2O_3$ in the final product. The nitrate salts were seperately dissolved in ethylene glycol by stirring for 45 min separately. The resulting solutions were then mixed thoroughly and stirred while heating till it formed a gel. The gel was then reduced in hydrogen atmosphere in an electric furnace. The reduction was carried out at different temperatures for different durations to get various sizes of the cobalt and chromium oxide particles as detailed in Table.I. The nanocomposites so formed were characterized by X-ray diffraction (XRD) [Panalytical XPert Pro MPD] using Cu Kα radiation. A JEOL 2100 model transmission electron microscope (TEM) was used for imaging nanoparticles dispersion on a carbon coated copper grid. Two different compositions of FM Co and AFM $Cr_2O_3$ were studied. The samples with Co to $Cr_2O_3$ weight ratio of 30:70 are termed as Series I, while that with Co to $Cr_2O_3$ weight ratio of 40:60 are in Series II. Samples A, B and C belong to series I while Samples D and E fall in series II.

The magnetization measurements were carried out using SQUID magnetometer (Quantum Design, MPMS). The field cooled (FC) hysteresis loops were measured after cooling the sample in the presence of an external magnetic field of 30 kOe. Before applying the cooling field, the sample was heated to 350K. The zero field cooled (ZFC) hysteresis loops were measured after cooling the sample from 350K to the measuring temperature (50K and 300K) in absence of any external field. In both FC and ZFC cases the magnetic hysteresis loop was measured up to a maximum field of 20 kOe. For FC and ZFC magnetization versus temperature measurements, the cooling of the samples (in case of FC) and magnetization measurement were performed in presence of 100 Oe magnetic field.

**III. Results and Discussions:**

The XRD patterns obtained for different samples are shown in Figure.1. The patterns show the formation of cubic Co and rhombohedral $Cr_2O_3$ phases separately. The crystallite size increases with the increase in temperature and/or duration of heat treatment during reduction process as observed from the peak broadening in Fig.1. The crystallite size was calculated from Scherrer's formula [15] for Co and $Cr_2O_3$. The values are presented in Table.1.

The HRTEM images of samples A and C, shown in Figure.2, confirm the formation of nearly spherical nanoparticles. However, it is not clear from the images, if Co-$Cr_2O_3$ is forming a core-shell system. Hence we assume the existence of two structures, first the core-shell structure and second being Co nanoparticles embedded in a uniform $Cr_2O_3$ matrix. As both are equally probable, the sample can be a mixture of these two structures.

The FC and ZFC hysteresis loops of samples A, B, C, D and E are shown in Figures 3(a), 3(b), 3(c), 3(d) and 3(e) respectively at temperatures 50K and 300K. The samples show non-zero hysteresis loss at both temperatures. The existence of exchange bias between ferromagnetic Co and antiferromagnetic $Cr_2O_3$ systems can be verified by the shift of hysteresis loop along field axis and/or coercivity enhancement in FC configuration. The origin of this loop shift lies in the exchange interaction present between the interface spins of FM and AFM phase. In all the five cases this kind of loop shift is observed at 50K in FC configuration (Figure.3). However, a small exchange bias field is observed in ZFC measurements also, which is attributed to the effect of FM magnetic moment in aligning the AFM moments [16, 17]. This requires a considerable particle moment, which is possible in samples containing larger particles of Co. The amount of shift in hysteresis loop termed as exchange bias field ($H_{EB}$) is determined and presented in Table.1 along with the coercivity field $H_C$. Maximum $H_{EB}$ of 427 Oe is observed in case of sample C. However, the value of $H_{EB}$ in the present case is less compared with the corresponding values usually observed in multi-layer systems but in accordance with the values observed in powder samples [17]. At 50K, the enhancement of $H_C$ value is observed in FC case (except in case of Sample B) compared to that for ZFC samples. Both the shifting of loop and enhancement of $H_C$ indicate the presence of exchange bias between the Co and $Cr_2O_3$ interfaces. Samples B and D are of special interest as these samples consist of Co particles whose $T_B$ is well below room temperature [18]. But our measurements show that these samples are not superparamagnetic even at 300K. So the superparamagnetic effect in fine Co particles is overcome using exchange bias effect. The observation is similar to that reported by Skumryev et al [9] as mentioned before. We have made use of the fact that Neel temperature $T_N$ of $Cr_2O_3$ is higher (~310K for bulk $Cr_2O_3$) than of CoO (~290K).

From the FC and ZFC magnetization (M) Vs temperature(T) curves shown in Figure.4 it is clear that the blocking temperature is around or above 350K. Also a small value of $H_{EB}$ is observed at 300K. In exchange biased systems the blocking temperature of ferromagnetic phase is influenced by the Neel temperature of antiferromagnetic phase. Enhancement of Neel temperature of $Cr_2O_3$ with decrease in particle size has been reported earlier [19]. Hence the observed higher $T_B$ of the samples is in accordance with the $T_N$ value for $Cr_2O_3$. However, the nature of M Vs T may also suggest the existence of interparticle interactions and surface effects in small particles. We would like to verify this in the future. Nevertheless the hysteresis loop shift and/or coercivity enhancement prove the exchange bias effect in these samples and we suppose it is the reason for thermal stability of magnetic moment in ultra fine Co nanoparticles (Samples B and D) studied.

Though various efforts were made earlier to enhance the blocking temperature using exchange bias [20, 21], the present work deals with the lower particle sizes of a few nm and yet provides a mechanism for obtaining a $T_B$ above room temperature. In addition, the method used for sample preparation is very simple and cost effective. Hence the present system can be a competent possibility for the application in magnetic data storage systems. However, the system has to be synthesized in layered form and the possibility of increasing cobalt percentage is to be checked in order to attain higher data storage densities.

## IV. Conclusions:

We have demonstrated the exchange bias effect in Co-$Cr_2O_3$ nanocomposite powders, prepared by a simple chemical method. The blocking temperature of ~3nm Co nanoparticles is observed to be increased due to exchange bias effect. Hence this mechanism may be a possibility to overcome the superparamagnetic effect in small nanoparticles.

## Acknowledgements:

This work was supported by BRNS, Govt. of India under the project grant 2003/37/13/BRNS. One of the authors (PAK) thanks CSIR, Govt. of India for the research fellowship. We also acknowledge the help of Prof R Maitra and Mr R Bosu in TEM imaging.

## Figure captions:

**Figure.1:** X-ray diffraction pattern for various samples, showing the presence of Co ((111) peak) & $Cr_2O_3$ (remaining peaks) phases separately.

**Figure.2:** HRTEM micrographs of samples A, B and C. (Dark areas represent Co while lighter areas represent $Cr_2O_3$).

**Figure.3:** Partial FC-ZFC hysteresis loops of

**a)** Sample A at 50K (above) and 300K (below)

**b)** Sample B at 50K (above) and 300K (below)

**c)** Sample C at 50K (above) and 300K (below)

**d)** Sample D at 50K (above) and 300K (below)

**e)** Sample E at 50K (above) and 300K (below)

**Figure.4:** FC (closed circles) and ZFC (open circles), M Vs T curves of samples A, B and D. Field cooling and magnetization measurements were performed in presence of 100 Oe magnetic field.

**Table 1: Crystallite size, $H_{EB}$, $H_C$(300K), $H_C$(50K) for different samples.**

| Sample | Heat treatment $H_2$ | Crystallite size of Co from XRD (St. Dev.) | Crystallite size of $Cr_2O_3$ from XRD (St. Dev.) | $H_{EB}$ at 50K) (Oe) | $H_C$ at 300K (Oe) | | $H_C$ at 50K (Oe) | |
|---|---|---|---|---|---|---|---|---|
| | | | | | FC | ZFC | FC | ZFC |
| A Series-I | 500°C (3hrs) | 43nm (3.5) | 27.3nm (2.9) | 110 | 192 | 184 | 249 | 209 |
| B Series-I | 700°C (15min) | 3.1nm (0.4) | 3.2nm (0.5) | 85 | 296 | 270 | 355 | 502 |
| C Series-I | 700°C (2hrs) | 15.4nm (2.4) | 35.3nm (3.0) | 427 | 332 | 324 | 1037 | 934 |
| D Series-II | 600°C (1hr) | 3.3nm (0.6) | 4.4nm (0.7) | 150 | 98 | 104 | 315 | 279 |
| E Series-II | 700°C (3hrs) | 20.85 (2.7) | 24.40 (2.8) | 194 | 250 | 242 | 349 | 252 |

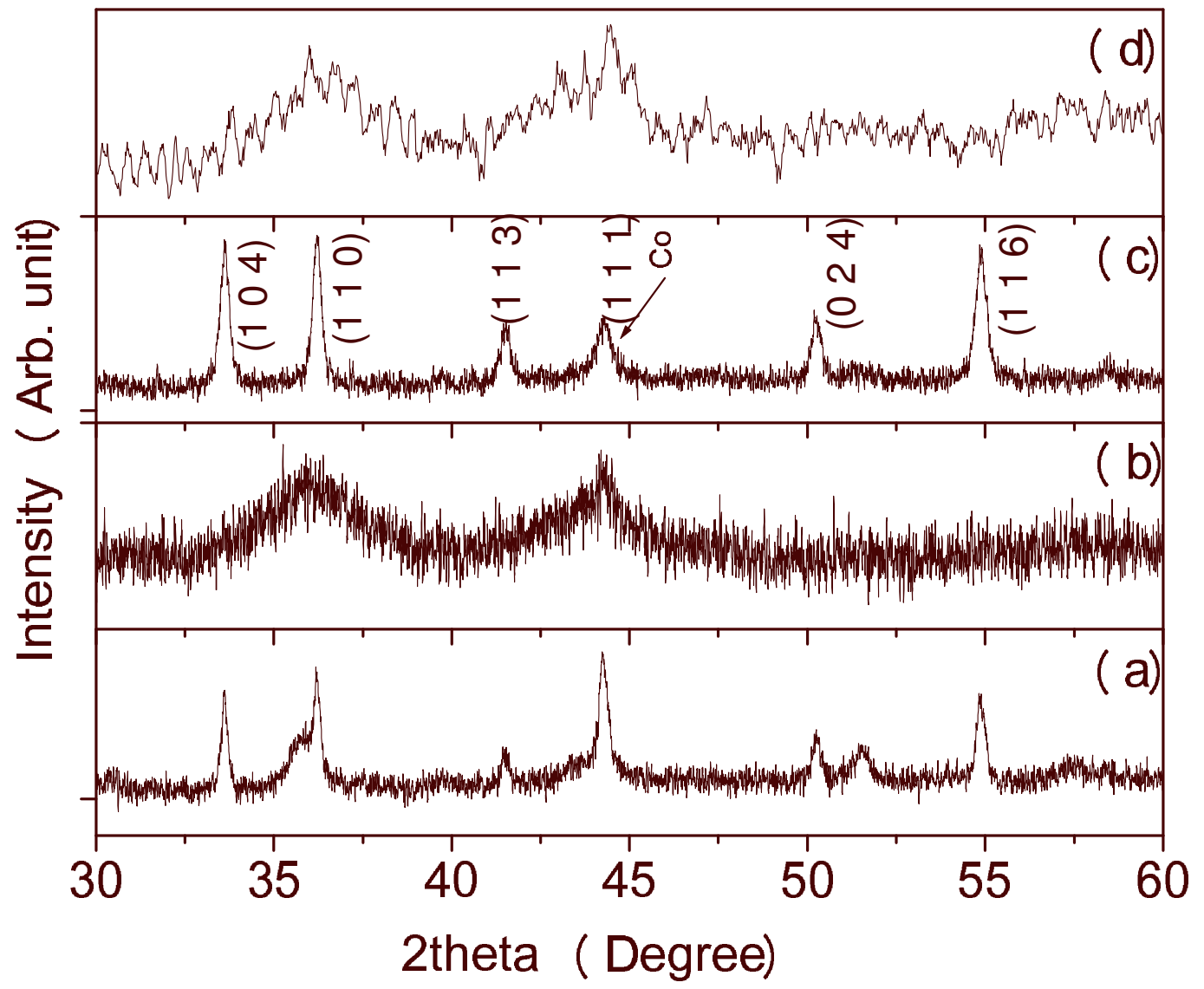

**Figure.1**

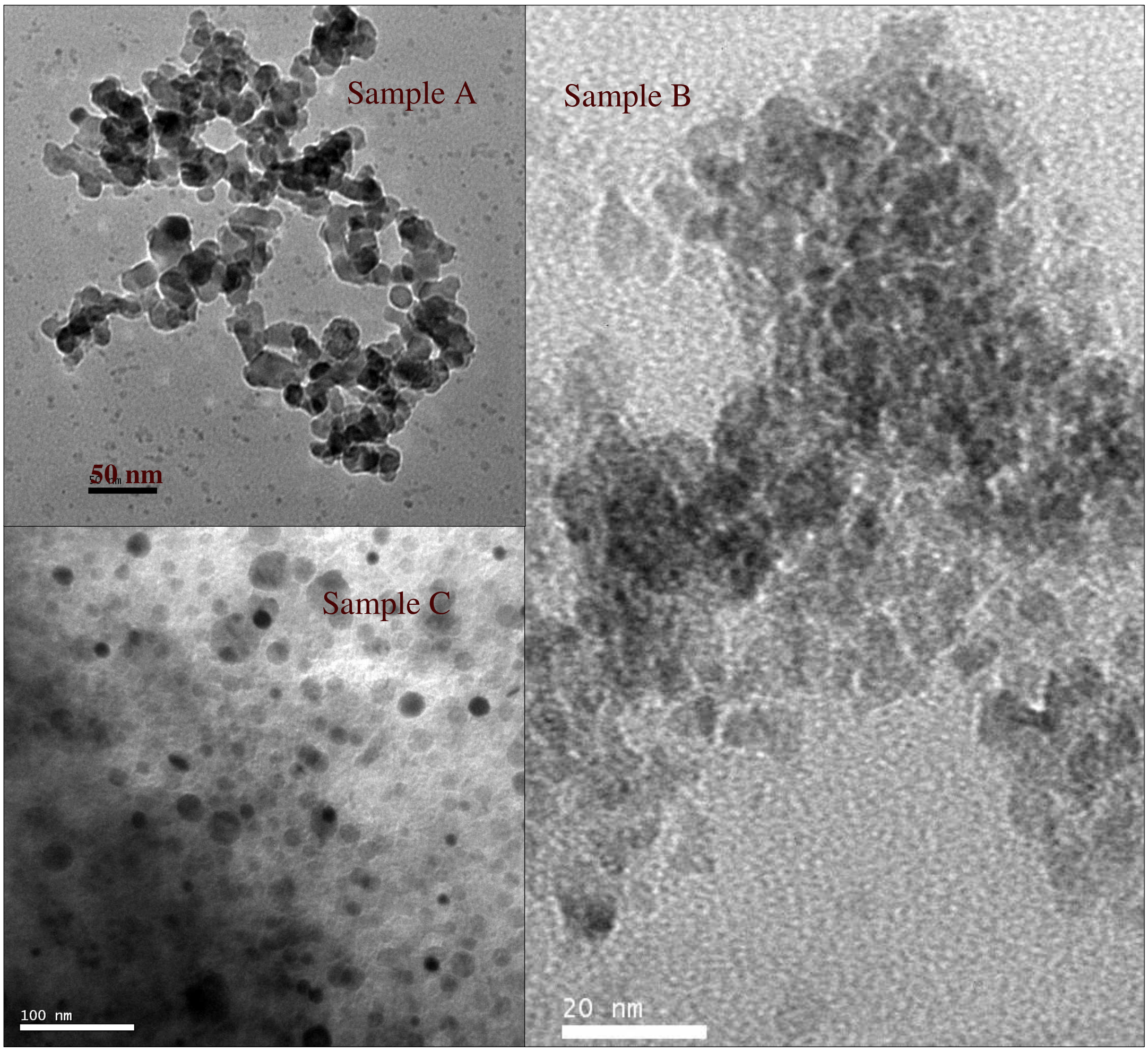

**Figure.2**

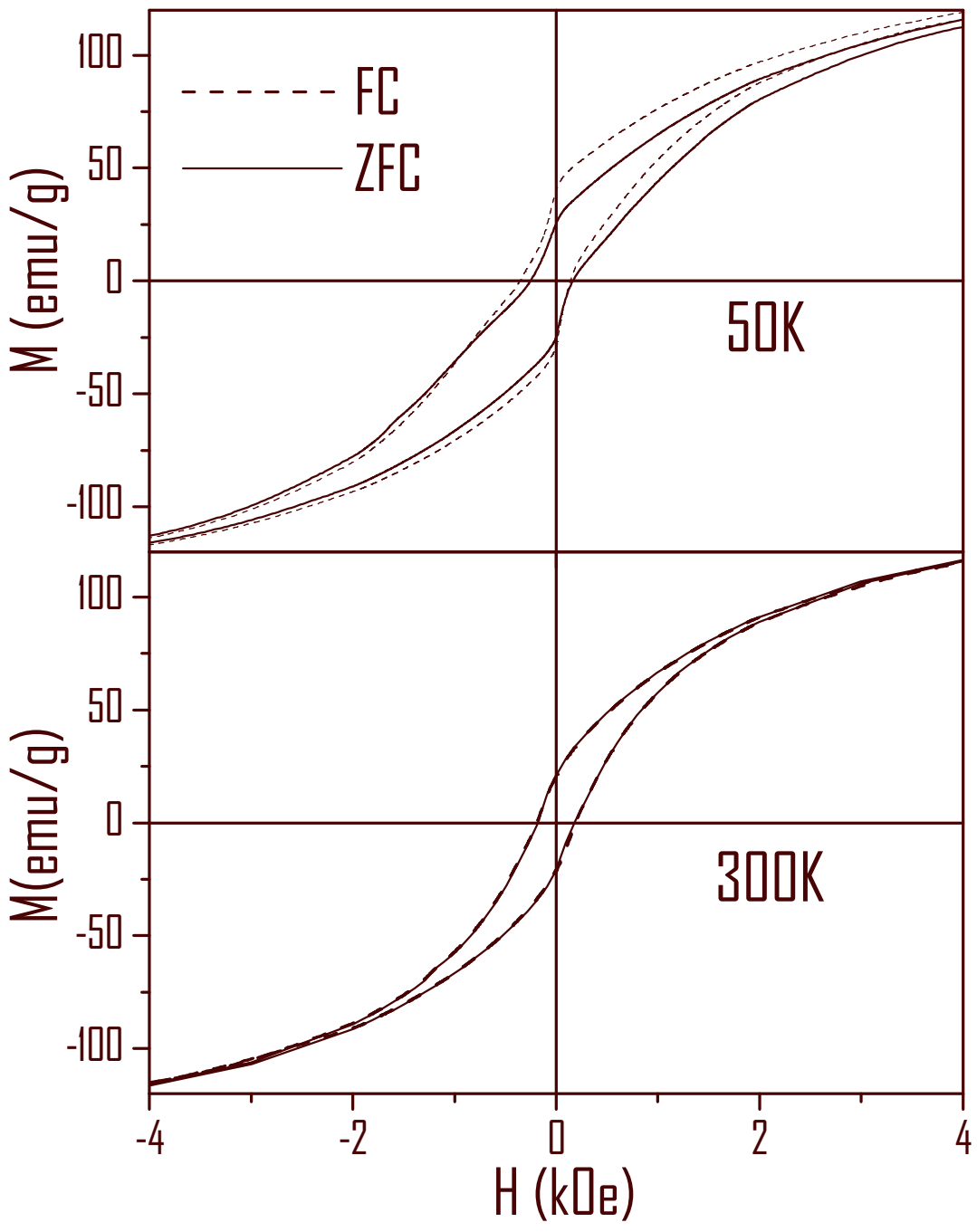

**Figure.3a**

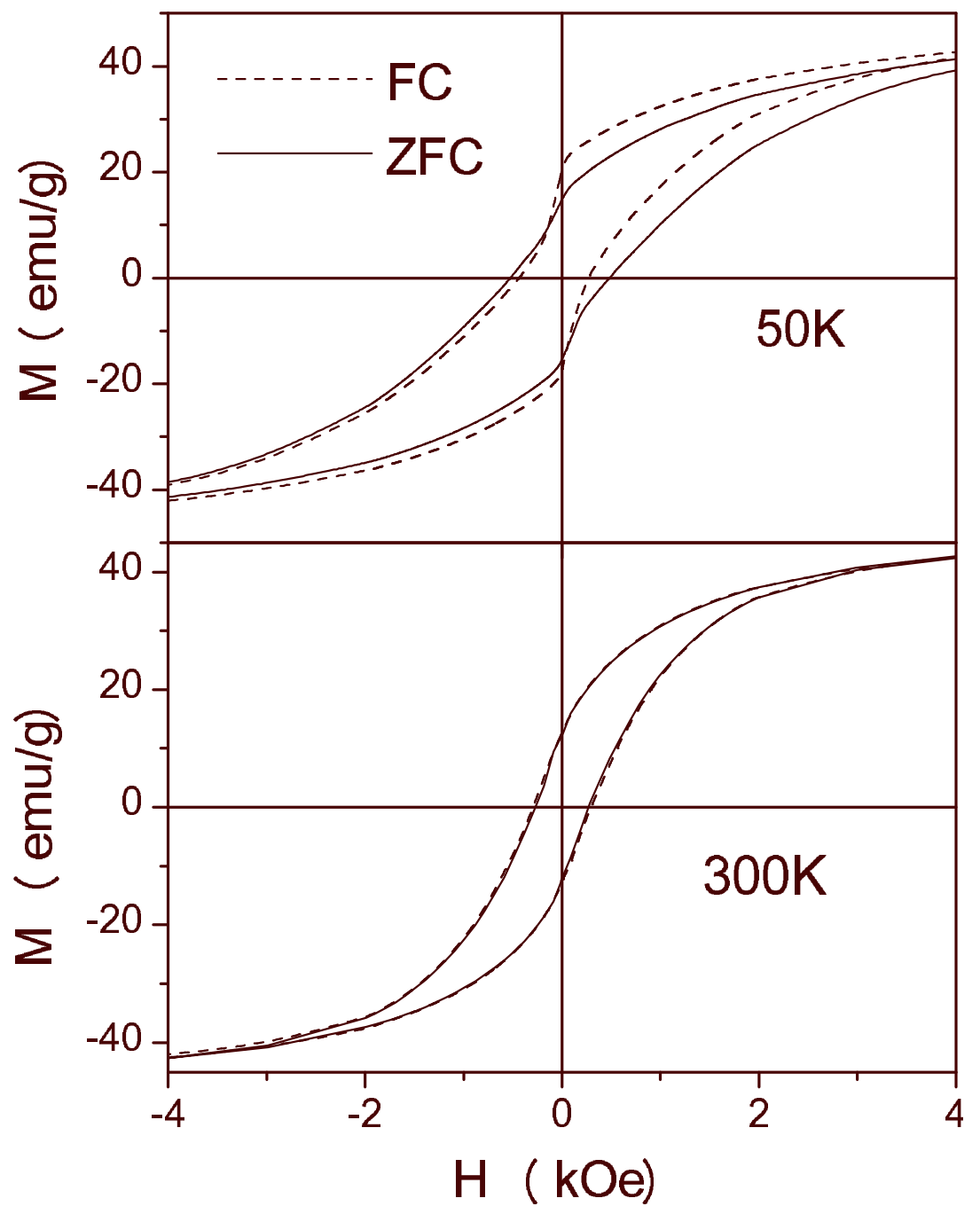

**Figure 3b**

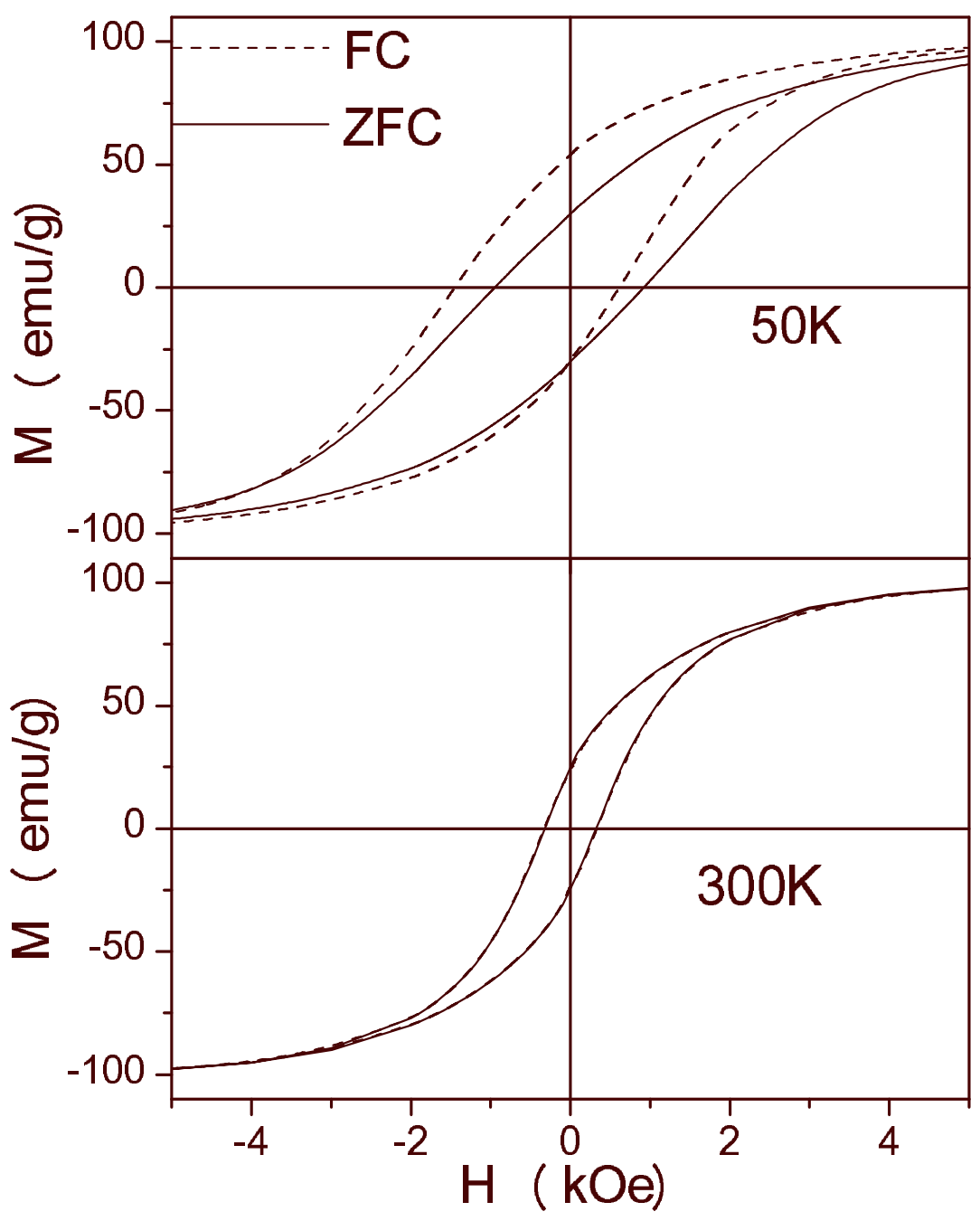

**Figure 3c**

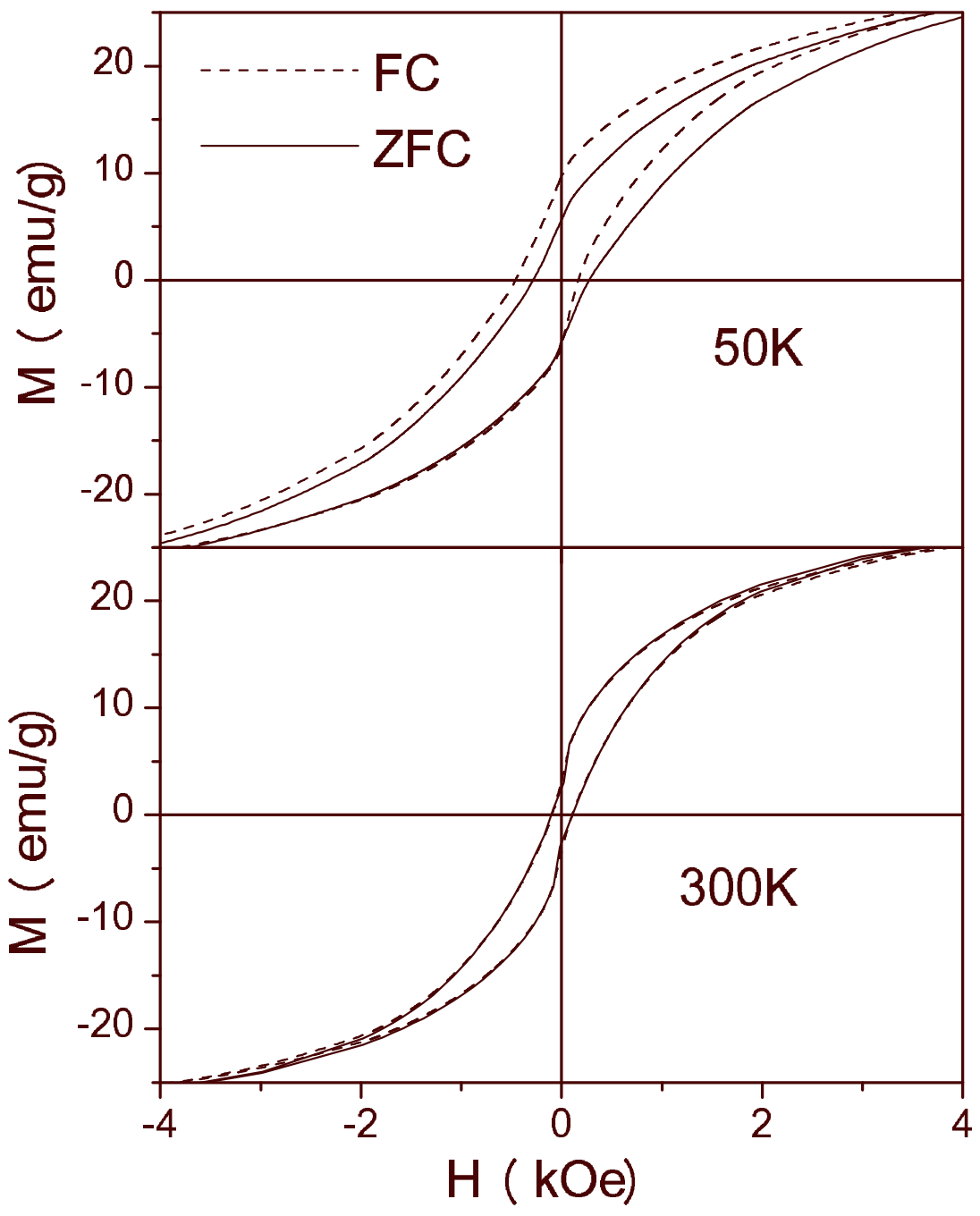

**Figure 3d**

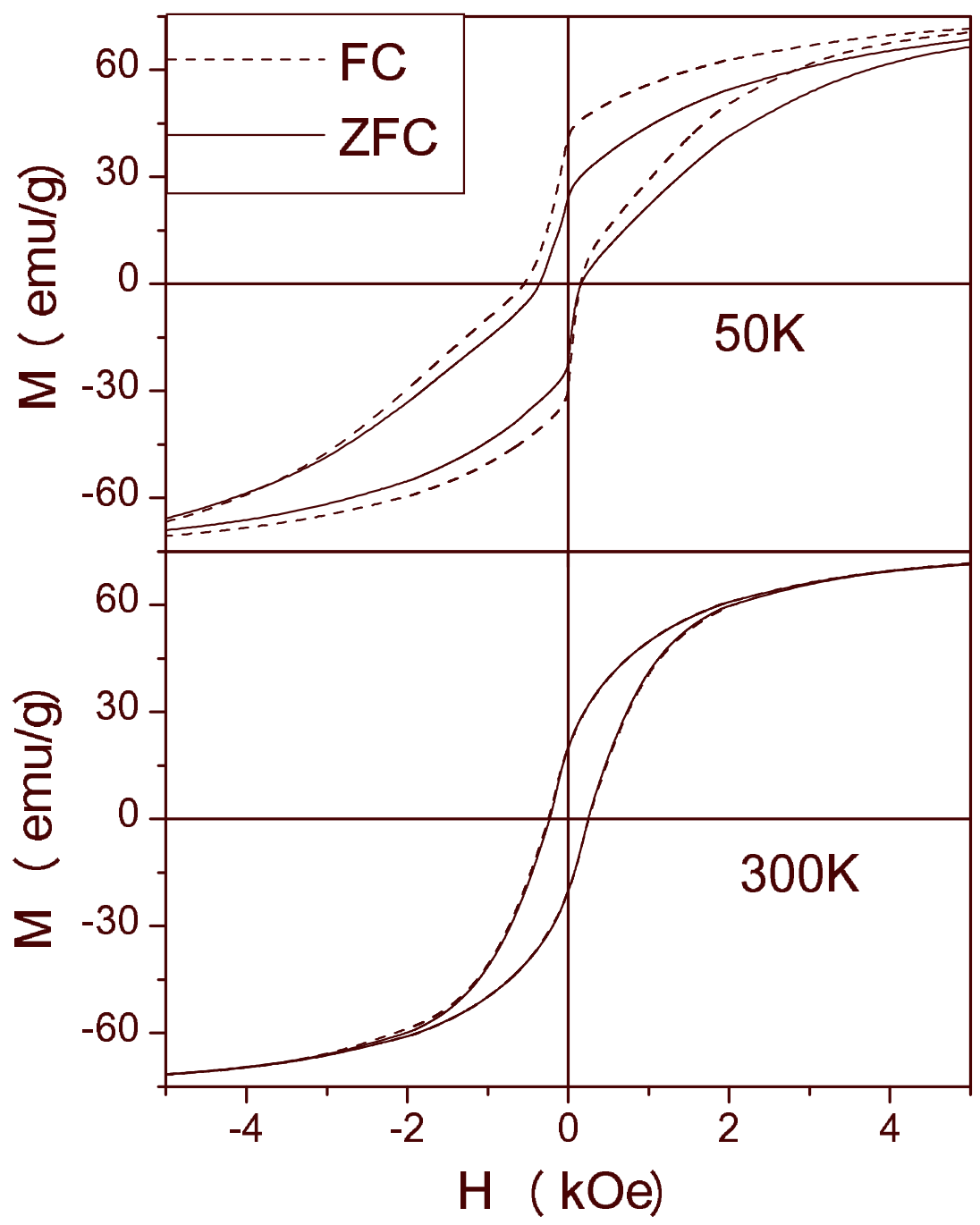

**Figure 3e**

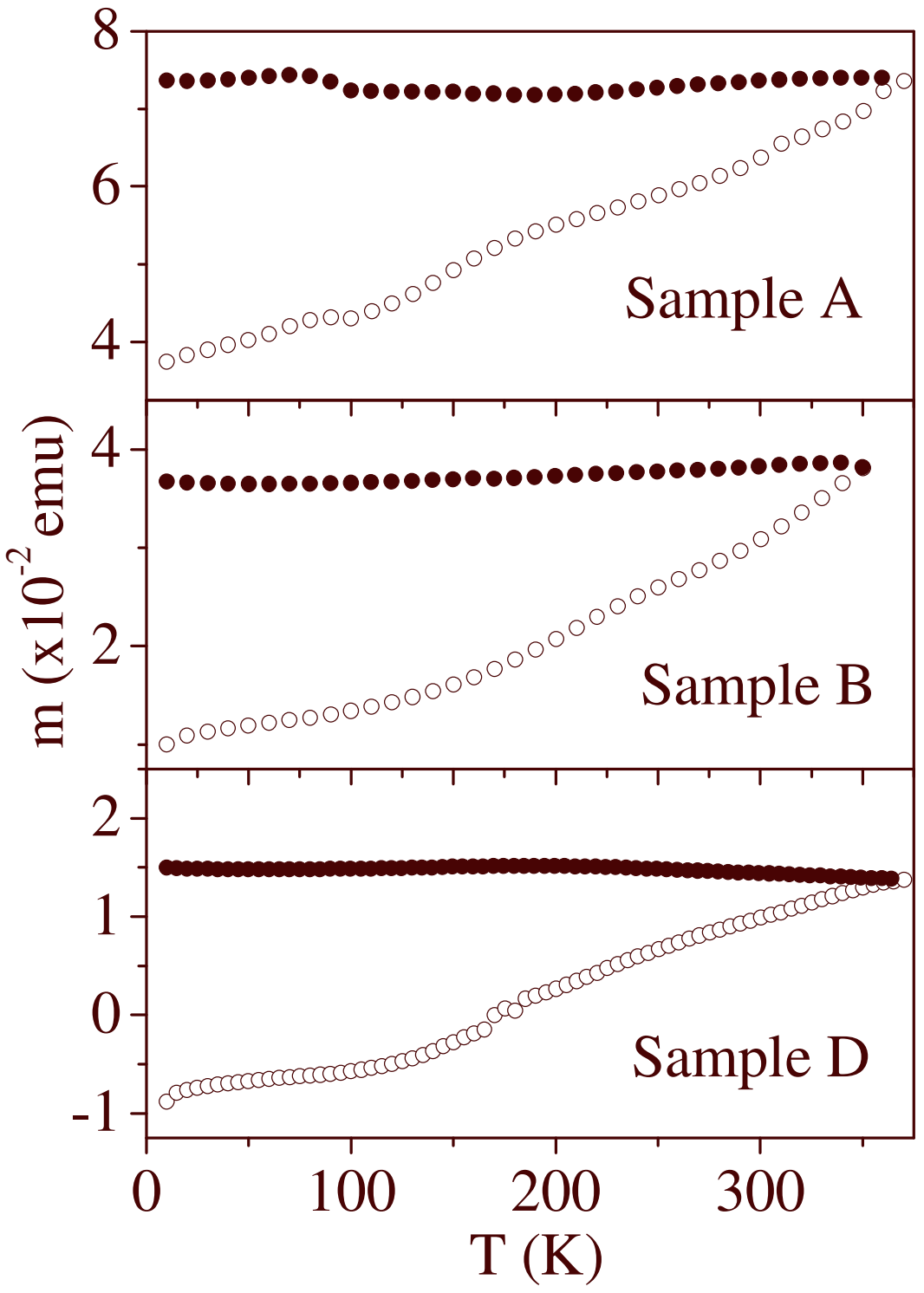

**Figure.4**